\documentstyle{article}
\def\abstract#1{\vskip 7mm
        \begin{center}{}\par \smallskip
                \begin{minipage}[c]{12cm}
                        \small #1
                \end{minipage}
        \end{center}
}
\def\title#1{\begin{flushleft}{\Large\bf #1}\end{flushleft}}
\def\author#1{\vskip 5mm \begin{flushleft}{#1}\end{flushleft}}
\def\address#1{\begin{flushleft}{\it #1}\end{flushleft}}
\begin{document}

\setlength\arraycolsep{2pt}

\title{Cosmological Implications of a Non-Separable 5D Solution\\
of the Vacuum Einstein Field Equations}

\author{Takao Fukui\footnote{E-mail: \tt{fukui@sciborg.uwaterloo.ca}}%
${}^,$\footnote{Permanent address: Dokkyo University, Soka,
Saitama 340-0042, Japan}, Sanjeev S.~Seahra\footnote{E-mail:
\tt{ssseahra@sciborg.uwaterloo.ca}} and Paul
S.~Wesson\footnote{E-mail: \tt{wesson@astro.uwaterloo.ca}}}

\address{Department of Physics, University of Waterloo\\
Waterloo, Ontario, N2L 3G1, Canada}

\abstract{An exact class of solutions of the 5D vacuum Einstein field
equations (EFEs) is obtained.  The metric coefficients are found
to be non-separable functions of time and the extra coordinate
$l$ and the induced metric on $l$ = constant hypersurfaces has
the form of a Friedmann-Robertson-Walker cosmology.  The 5D
manifold and 3D and 4D submanifolds are in general curved, which
distinguishes this solution from previous ones in the literature.
The singularity structure of the manifold is explored: some
models in the class do not exhibit a big bang, while other
exhibit a big bang and a big crunch.  For the models with an
initial singularity, the equation of state of the induced matter
evolves from radiation like at early epochs to Milne-like at late
times and the big bang manifests itself as a singular
hypersurface in 5D. The projection of comoving 5D null geodesics
onto the 4D submanifold is shown to be compatible with standard
4D comoving trajectories, while the expansion of 5D null
congruences is shown to be in line with conventional notions of
the Hubble expansion.
}

\section*{I.~INTRODUCTION}
The vacuum EFEs for spacetime plus an extra dimension are
given in terms of the Ricci tensor by $R_{AB}=0 \,\, (A,B=0-3,4)$.
These contain the field equations of general relativity, which in
terms of the Einstein tensor and an induced energy-momentum tensor
are $G_{\alpha\beta}=8\pi T_{\alpha\beta} \,\, (\alpha,\beta
=0-3)$. The latter is obtained by a well-known technique.$^1$
Mathematically it can depend on $g_{4\alpha}$, $g_{44}$ and
derivatives of $g_{\alpha\beta}$ with respect to $x^4=l$, while
physically it can describe a perfect fluid with density $\rho$ and
pressure $p$. If there is no dependency on $l$, the equation of
state is that of radiation or ultrarelativistic particles, $p=\rho
/3$.$^2$  If there is dependency on $l$, a wide range is available
for the equation of state.$^3$ What are often referred to as the
standard 5D cosmological models were found as a class of solutions
of $R_{AB}=0$ by Ponce de Leon.$^4$ These solutions are separable
in the time $(t)$, space $(r,\theta ,\phi)$ and the extra
coordinate $(l)$. On the hypersurfaces $l=$ constant which we
label $\Sigma_l$, they reduce to the standard
Friedmann-Robertson-Walker (FRW) models of 4D cosmology with flat
space sections $(k=0)$. The class depends on one dimensionless
parameter which fixes the scale-factor $a$ for the dynamics and
the equation of state for the matter. It includes the $a=t^{2/3}$,
$p=0$ Einstein-de Sitter solution for the late universe, and the
$a=t^{1/2}$, $ p=\rho /3$ radiation solution for the early
universe. However, the Ponce de Leon solutions are not unique as
the 5D analysis of the standard 4D FRW solutions.

In Section II, we obtain a non-separable solution of the scale
factor. Sections III and IV are devoted to cosmological implications,
i.e.~singularities and geodesics respectively. Section V is for
comments.

\section*{II.~NON-SEPARABLE SOLUTION}

A class of solutions which is mathematically different and
physically reasonable has a 5D line element given by
\begin{equation}
dS^2=\sigma
\left\{dt^2-a^2\left[\frac{dr^2}{(1-kr^2)}+r^2d\Omega^2\right]\right\}+\epsilon
b^2dl^2.
\end{equation}
Here $k=\pm1,0$ describes the 3D curvature, $d\Omega^2\equiv
d\theta^2+\sin^2\theta \, d\phi^2$ describes the 2D spherical
geometry, and $\epsilon=\pm1$ describes the nature of the extra
dimension. The functions $\sigma=\sigma(l), ~a=a(t,l), ~b=b(t,l)$
are to be fixed by the 5D field equations. We have studied the
latter, particularly with regard to the 4D properties of matter.
Hereafter we employ $c=G=1$ unit system, that is, $L=M=T$.

A simple class of solutions of $G_{AB}=0$ is obtained if we write
$\sigma= \mathrm{constant} \equiv\sigma_0$, $\hat{a}\equiv\partial
a/\partial l=f(l)b$ (here and henceforth, we use hats to denote
$\partial / \partial l$). Then the scale factors for ordinary
spacetime and the extra dimension are
\begin{equation}
    a=\sqrt{-Ft^2+gt+h},
\end{equation}
\begin{equation}
    b=\frac{-2\epsilon\sigma_0 f\hat{f}t^2+\hat{g}t+\hat{h}}{2fa}.
\end{equation}
Here $F\equiv \epsilon\sigma_0 f^2+k$, $g=g(l)$ and $h=h(l)$ are
functions, so including $f=f(l)$ we have three such. There exists
are relation between these, set by the field equations. It is
\begin{equation}
    h=-\frac{g^2+\kappa}{4F},
\end{equation}
where $\kappa$ is a constant with the physical dimensions of
(length)$^2$. In terms of this, the 3D scale factor (2) is given
by
\begin{equation}
    a^2=-\frac{[(2Ft-g)^2+\kappa]}{4F}.
\end{equation}
The extra-dimensional scale-factor (3) is given by
\begin{equation}
    b=-\frac{[4\epsilon\sigma_0 f \hat{f}F
    t^2-2\hat{g}F^2t+g\hat{g}F-\epsilon\sigma_0
    f\hat{f}(g^2+\kappa)]}{4afF^2}.
\end{equation}
The new solutions have some properties in common with the Ponce de
Leon cosmologies$^4$ and some which are different. The Ponce de
Leon solutions only exist for $\epsilon=-1$, but the new ones have
$\epsilon=\pm1$ so the extra dimension can be spacelike or
timelike. Other (wave-like) solutions are known for $\epsilon=+1$,$^5$
 but the new solutions open the way to testing the signature
of the 5D manifold by observations of the 4D properties of matter
as given by (8) and (9) below. The solution (6) is in general
somewhat complicated, but we will not be much concerned with it
because it is (5) which, via the Friedmann equations, determines
the properties of matter. These for perfect fluid mean that
Einstein's equations read as usual
\begin{equation}
G_{\alpha\beta}=8\pi T_{\alpha\beta}=8\pi[(\rho+p)u_\alpha u_\beta
-pg_{\alpha\beta}].
\end{equation}
Here the 4-velocities $u^\alpha\equiv dx^\alpha/ds$ are defined in
terms of the 4D interval $s$,  which is included in $S$ of (1),
and we have $u^\alpha=(1,0,0,0)$. Then the density and pressure
are given by
\begin{equation}
\frac{8\pi\rho}{3}=-\frac{\epsilon
f^2}{a^2}-\frac{\kappa}{4\sigma_0a^4},
\end{equation}
\begin{equation}
8\pi p=\frac{\epsilon f^2}{a^2}-\frac{\kappa}{4\sigma_0a^4}.
\end{equation}
We recover FRW-like models on  $\Sigma_l$ that contain an exotic
type of induced matter, with an equation of state which follows
from (8) and (9),
\begin{equation}
p=\frac{\rho}{3}\left[1-\frac{8\epsilon\sigma_0
a^2f^2}{(4\epsilon\sigma_0 a^2f^2+\kappa)}\right].
\end{equation}
This equation of state is manifestly dependent on time and
completes the formal part of our analysis, which a fast computer
package$^6$ has confirmed. The form of these relations suggests a
two-fluid model as used in solutions of straight general
relativity. The first terms in (8), (9) by themselves imply $(\rho
+3p)=0$, which is the 4D signature of matter with zero
gravitational density.$^{1,7}$ This kind of matter has been
suggested as relevant to cosmic strings, i.e. open strings with sub-relativity transverse motions,$^8$ and K-matter which existence mentions the possibility of a universe dominated by cosmic strings,$^9$ to zero-point fields required by quantum theory$^{10,11}$ and to extreme sources for the Reissner-Nordstr\"{o}m metric which require that the material contents of the sphere has no effect on gravitational interactions at its centre.$^{12}$ The second terms in (8), (9) by themselves imply $p=\rho
/3$, which of course means photon-like matter. These
identifications are supported by the behaviour of the scale-factor
(2) or (5), where the latter may vary as $t^{1/2}$ in the
radiation model or as $t$ in the empty (Milne) model. They become
exact when in (8), (9) $f\rightarrow 0$, $\kappa\rightarrow 0$
respectively. Or on a given $\Sigma_l$ hypersurface, we find in
(2) that as t$\rightarrow \infty$, $a(t,l)\rightarrow \infty$ ---
provided that $F<0$ --- which implies in (10) that $p\rightarrow
-\rho/3$. This is the equation of state of the Milne, or empty
universe where the gravitational density of matter is zero. Hence,
it might be understood that the 4D cosmologies on the $\Sigma_l$
hypersurface are asymptotically empty as $t\rightarrow \infty$.
However, it should be noted that $\rho$, $p$ of (8), (9) refer to
the total density and pressure, and that any split is in general
arbitrary in the absence of information about the particles which
make up the matter. Some of the latter may, in principle, be of
non-standard type, since $g_{44}=\epsilon b^2(t,l)$ describes a
3D-homogeneous scalar-field.$^1$ This can manifest itself in 4D as
a time-dependent cosmological ``constant", as required to
harmonize astrophysical data on the age of the universe.$^{13,14}$  A
time-dependent scalar field generalizes the constant vacuum of
Einstein's theory, whose density and pressure are given in terms
of the cosmological constant by $\rho = -p = \Lambda/8\pi$. The positive $\Lambda$ is concluded by revisiting the Tinsley diagram with the recent determinations of the Hubble constants.$^{15}$

\section*{III.~SINGULARITIES}
The age of the universe is defined in 4D models as the time
elapsed since the big bang, but the latter concept has to be
treated carefully in 5D models. In solutions of the 5D field
equations which are flat in 5D but contain a curved space which is
singular in 4D, the big bang has to be identified as a defect of
the geometry.$^{1,4}$

To analyze the early behaviour of the 4D cosmologies embedded in
(1), we need to establish whether or not these models contain a
big bang singularities. Related to this issue is the nature of the
singularity structure of 5D manifold as related to the singularity
structure of the 4D sub-manifolds. It is well known that the 5D
Ponce de Leon cosmological metrics contain 4D $\Sigma_l$
hypersurfaces that precisely mimic standard FRW cosmologies
complete with a big bang singularity. However, the 5D manifold is
flat $R^A{}_{BCD}=0$, which suggests that the 4D big bang is
merely a consequence of the way in which the $\Sigma_l$
hypersurfaces are embedded in 5D Minkowski space, or,
equivalently, the choice of 5D coordinates. The Ponce de Leon
solutions are flat in 3D, curved in 4D and (perhaps surprisingly)
flat in 5D.$^{1,16}$ That is, the Riemann-Christoffel tensors for 5D
and 4D are $R^A{}_{BCD}=0$ and $R^\alpha{}_{\beta\gamma\delta}\neq
0$, so a flat manifold smoothly embeds a curved one (locally),
like 3D Euclidean space embeds the 2D surface of the Earth. We
wish to address the issue of whether or not the present manifold
(1) is curved or flat, and whether or not it contains a genuine
singularity.

A direct attack on the problem could entail the calculation of the
5D Kretschmann curvature invariant $K\equiv R^{ABCD}R_{ABCD}$ for
(1). The divergence of this quantity is usually interpreted as
being indicative of a curvature sigularity in the manifold.$^{1,17}$
However, the calculation of $K$ for (1) is difficult, even by
computer. We can make some headway by considering a special case
defined by:
\begin{equation}
\epsilon=-1, ~k=0, ~f(l)=1/\sqrt{2\sigma_0}, ~g(l)=l.
\end{equation}
In this model we have
\begin{equation}
a(t,l)=\sqrt{\frac{(t+l)^2+\kappa}{2}}.
\end{equation}
The associated curvature invariant is
\begin{equation}
K=\frac{72\kappa^2}{\sigma_0^2[(t+l)^2+\kappa]^4}.
\end{equation}
The scale factor $a(t,l)$ goes to zero and the Kretschmann scalar
$K$ becomes infinity along the hypersurfaces
\begin{equation}
0=t+l\pm \sqrt{-\kappa}.
\end{equation}
Clearly, there will be no curvature singularity for $\kappa>0$.
Therefore, at least one specific case, we find that the manifold
(1) is curved in 5D and singular where the scale factor $a(t,l)$
vanishes, in sharp contrast with the Ponce de Leon solutions. For
the present model, the density and pressure of (8) and (9) will
diverge in general as $a\rightarrow 0$ along the hypersurfaces of
(14).

It is interesting to note that $K=0$ for $\kappa=0$ in the model
presented above. Indeed, an analysis of the general metric (1) via
computer shows that $R^A{}_{BCD}=0$ for $\kappa =0$. In this
eventuality, the scale factor (2) with (4) becomes
\begin{equation}
a(t,l)=\frac{g(l)t+2h(l)}{2h^{1/2}(l)},
\end{equation}
which is linear in time, just like the scale factor for the Milne
universe (indeed, the equation of state of the induced matter is
$p=-\rho/3)$. So the $\kappa=0$ case entails induced matter with
zero gravitational density, is curved in 4D and is flat in 5D.
However, it bears repeating that when $\kappa\neq 0$, it is
possible to have curved 5D solutions.

Now, by analogy with the standard FRW model, it is obvious that
the 4D sub-manifolds on $\Sigma_l$ will be singular for times
$t_{\ast}(l)$ such that $a(t_{\ast},l)=0$, an epoch commonly
referred as the big bang. Solving $a(t_{\ast},l)=0$ in (5) for $t_{\ast}$ gives two
solutions:
\begin{eqnarray}
t_{\ast}^{\pm}(l)=\frac{g(l)\pm\sqrt{-\kappa}}{2F}.
\end{eqnarray}
Two things are apparent from this result. First, there can be no
4D big bang if $\kappa>0$. This is in agreement with the special
case presented above, where the scale factor can only vanish if
$\kappa \leq 0$. Second, we see that if $\kappa<0$ there are two
separate singularities, or, in other words, two big bangs in $F<0$. These
4D events are located on the two 4D hypersurfaces defined by (16).
In the neighborhood of the big bang, $a^2(t_{\ast}^{\pm}+\delta
t,l)\sim a_{\ast}(l)\delta t$ where $a_{\ast}(l)$ is some function
of $l$ at $t=t_{\ast}^{\pm}$. The equation of state of the
cosmological matter is then
\begin{equation}
p=\frac{\rho}{3}\left[\frac{\kappa -\tilde{a}_{\ast}(l)\delta
t}{\kappa + \tilde{a}_{\ast}(l)\delta
t}\right]=\frac{\rho}{3}-O(\delta t),
\end{equation}
where $\tilde{a}_{\ast}(l)$ is some other function of $l$. As
$\delta t\rightarrow 0 $, we recover $p=\rho/3$. So, near the big
bang(s), the induced matter behaves like radiation while at late
times it behaves like the matter in the Milne model.

Finally, we consider the case $F>0$, which is guaranteed for
$\epsilon =1$ and $k=0,1$ since we must have $\sigma_0>0$. If we
also have $\kappa<0$, it is clear that on $\Sigma_l$, $a(t,l)$ is
real only between $t_\ast^-$ and $t_\ast^+$. This is a
cosmological model with both a big bang and a big crunch, similar
to the standard $k=+1$ FRW metrics. There therefore exists a
hiearchy of 4D cosmologies on the $\Sigma_l$ hypersurfaces. If
$\kappa<0$, then the issue of whether each of the cosmologies is
forever expanding or destined to end in a big crunch is entirely
determined by the value of $f(l)$ on that hypersurface. If $
\kappa\geq 0$, the quantity inside the radical in the definition
of $a(t,l)$ in (5) will be less than or equal to zero for all
times, which makes $a(t,l)$ complex and has the effect of
switching  the signature of the $(x^1,x^2,x^3)$ coordinates from
spacelike to timelike (this also happens for $t\not\in
[t_\ast^-,t_\ast^+]$ if $\kappa<0$).  Table 1 summarizes all the
case discussed in this section for both the global 5D geometry and
the type of cosmology embedded on the $\Sigma_l$ hypersurfaces.\\

 \fbox{Table 1}

\section*{IV.~GEODESICS}

The possibility the test particles move on higher dimensional
geodesics has been studied by many authors.$^{1,18-20}$  In
particular, it has recently been demonstrated that particles that
are massless 5D appear to be massive in 4D.$^{21}$ In this section,
we examine the 5D null and comoving geodesics of the metric (1)
for  the $\epsilon=-1$ case and identify the 4D GR limit.

We take $\epsilon=-1$ and define $\Sigma^2\equiv\sigma_0$. Then,
the metric (1) may be written as
\begin{equation}
dS^2=[\Sigma dt+b(t,l)dl][\Sigma
dt-b(t,l)dl]-\Sigma^2a^2(t,l)ds_3^2,
\end{equation}
where $\Sigma^2a^2(t,l)ds_3^2$ represents the spatial 3-metric,
with
\begin{equation}
ds_3^2=\frac{dr^2}{1-kr^2}+r^2d\Omega^2.
\end{equation}
Now, since the terms in square brackets represent a two
dimensional manifold, which must be conformally flat, it is in
principle possible to find a coordinate transformation
\begin{equation}
\eta=\eta(t,l),\hspace{5mm} \xi=\xi(t,l)
\end{equation}
that will cast the metric in the form
\begin{equation}
dS^2=C(\eta,\xi)d\eta d\xi - \Sigma^2a^2(\eta, \xi)ds_3^2.
\end{equation}
Then, the null geodesics of the manifold that are spatially
comoving ($ds_3^2=0$) are just the $d\eta=0$ or $d\xi=0$
trajectories. However, the transformation (20) is not immediately
obvious. Therefore, we must be content with an indirect analysis
of the properties of the 5D null-comoving paths.

Consider the vectors
\begin{eqnarray}
k^A\partial_A=\frac{1}{\sqrt{2}\Sigma}\partial_t +
\frac{1}{\sqrt{2}b(t,l)}\partial_l,\\
N^A\partial_A=\frac{1}{\sqrt{2}\Sigma}\partial_t -
\frac{1}{\sqrt{2}b(t,l)}\partial_l.
\end{eqnarray}
They satisfy
\begin{equation}
k^A k_A = 0, \quad  N^A N_A = 0, \quad k^A N_A=1.
\end{equation}
Both $k^A$ and $N^A$ are tangent to null geodesics and can be
thought of as the principle and auxiliary vectors of a null
congruence. They both share the same parameterization, defined by
\begin{eqnarray}
\frac{dt}{d\lambda}&=&\frac{1}{\sqrt{2}\Sigma},\nonumber
\end{eqnarray}
\begin{eqnarray}
    \frac{dl}{d\lambda}& = & + \frac{1}{\sqrt{2}b(t,l)} \qquad
    ({\mathrm{for}}~k^A), \nonumber \\ &=&-\frac{1}{\sqrt{2}b(t,l)} \qquad
    ({\mathrm{for}}~N^A).
\end{eqnarray}
In order to determine if $\lambda$ is an 5D affine parameter, we
calculate
\begin{equation}
k^A\nabla_Ak^B=\left[\frac{1}{\sqrt{2}\Sigma}\frac{\partial}{\partial
t}\ln b(t,l)\right]k^B,
\end{equation}
with a similar expression for $N^A$. $\nabla_A$ is the 5D
covariant derivative operator. This shows that $\lambda$ is not a
5D affine parameter. However, it is easy to see that the 4D
projections of $k^A$ and $N^A$ onto $\Sigma_l$ satisfy the 4D
geodesic equation for timelike geodesics. Therefore, $\lambda$ is
the 4D proper time $\tau=\Sigma t$ comoving geodesics (up to a
constant prefactor). This is an excellent example of how a null
path in 5D can appear to be the trajectory of a massive particle
in 4D.

We now explore the possibility that galaxies travel along 5D
trajectories described by (22) or (23). The tangent space to the
null vectors $k^A$ and $N^A$ is necessarily three dimensional and
has the metric
\begin{equation}
q_{AB}=g_{AB}-k_AN_B-k_BN_A,
\end{equation}
which gives
\begin{equation}
q_{AB}dx^Adx^B=-\Sigma^2a^2(t,l)ds_3^2.
\end{equation}
It is clear that the tangent space of $k^A$ and $N^A$ is
equivalent to the $t=$ constant, $l=$ constant 3-surface of the 5D
manifold. Hence, observers traveling along 5D null-comoving
geodesics 4D comoving geodesics (confined to $\Sigma_l$) share the
same tangent space, which suggests that they will observe the
world around them in the same manner. In particular, we can
calculate the expansion $\Theta$ of a congruence of 5D geodesics
with tangent vector $k^A$. For non-affinity parameterized
geodesics, $\Theta$ is given by:
\begin{equation}
\Theta=\nabla^Ak_A-N_Bk^A\nabla_Ak^B.
\end{equation}
Plugging in expression for $k^A$ and $N^A$, we get
\begin{equation}
\Theta=\frac{3}{a}\left(\frac{1}{\sqrt{2}\Sigma}\frac{\partial
a}{\partial t}+\frac{1}{\sqrt{2}b(t,l)}\frac{\partial a}{\partial
l}\right)=\frac{3}{a}\frac{da}{d\lambda}.
\end{equation}
Since the expansion scalar represents the fractional rate of
change in the 3-volume $\delta V$ of the
congruence\footnote{i.e.~$\Theta=(\delta V)^{-1}d(\delta
V)/d\lambda$}, we see that $\delta V\sim a^3(\lambda)$ as
expected. That is, an observer in a galaxy traveling along a 5D
null-comoving geodesic will see the other galaxies receding away
in a Hubble-like expansion with scale factor $a(\lambda)$ (recall
that $\lambda=\sqrt{2}\tau$).

We are now in a position to identify the 4D limit of the theory.
The geodesic paths described by $k^A$ and $N^A$ would be identical
to 4D geodesics if $dl=0$, that is, they are confined to
$\Sigma_l$ hypersurfaces. This will be true if
$|dl/d\tau|=b^{-1}(t,l)\ll 1$, which can be physically interpreted
as demanding that large changes in the 4D proper time $\tau$ be
accompanied by small changes in the extra coordinate $l$. This can
be accomplished in the $\epsilon=-1$ case by choosing $f(l)$ and
$\sigma_0$ such that $|f(l)|\ll1$ and $[f^2(l)\sigma_0-k]>0$,
which ensures that $a(t,l)$ is real as $t\rightarrow\infty$. In
terms of the toy model (11), we recover the 4D limit for
$\Sigma\rightarrow \infty$.

\section*{V.~COMMENTS}

The new non-separable solution (2) gives a physical meaning of the
coefficients $K_1$  and $K_2$  which appear in the solution of the
scale factor.$^{22}$  That is, these constants and the coefficient of
$t^2$ as well are originated in $f(l), g(l)$ and $h(l)$ on the
hypersurface $\Sigma_l$. This helps us to consider the special case (11).

The physical big bang in 4D is a hypersurface in 5D. This kind of
behaviour has been observed in other 5D solutions, e.g. in a ``wave-like" class of exact cosmological solutions which look like waves propagating in the fifth dimension.$^{23}$ For the
example just considered, the 4D big bang is akin to a 5D shock
wave.$^{24}$ It is different to what happens in the Ponce de Leon
cosmologies,$^4$  because of the non-separable nature of the
solutions. In general, 5D cosmologies where the 4D big bang is a
geometrical effect have major implications for the early universe,
notably in regard to particle masses during the inflationary epoch$^{25}$ and the thermalization of photons by the particle mass varying quadratically with the time during the subsequent
radiation epoch.$^{26}$  5D cosmologies imply the variability of the masses of all particles.$^1$  The precise nature of the transformation
(20) for general choices of $f(l), ~g(l)$ and $h(l)$ will naturally
give us closed form solutions for the comoving null geodesic
equation and enable us to study those cosmological phenomena in
the early universe. These studies and a study of timelike 5D
geodesics ($\epsilon=1$) are left for the future work.
\section*{ACKNOWLEDGEMENTS}
  S.S.S.~would like to thank NSERC of Canada for financial support. T.F. is grateful for the hospitality of the Department of Physics, University of Waterloo while staying on the research program of Dokkyo University.

\section*{}
$^{~1}$~P.~S.~Wesson, {\it Space,Time,Matter}(World Scientific,Singapore,1999).\\
$^{~2}$~P.~S.~Wesson and J.~Ponce de Leon, {\it J.~Math.~Phys.} {\bf 33}, 3883 (1992).\\
$^{~3}$~J.~Ponce de Leon and P.~S.~Wesson, {\it J.~Math.~Phys.} {\bf34}, 4080 (1993).\\
$^{~4}$~J.~Ponce de Leon, {\it Gen.~Rel.~Grav.} {\bf 20}, 539 (1988).\\
$^{~5}$~A.~Billyard and P.~S.~Wesson, {\it Gen.~Rel.~Grav.} {\bf 28}, 129 (1996).\\
$^{~6}$~K.~Lake, P.~J.Musgrave and D.~Pollney, GR~Tensor(Dept. Phys.,Queen's University,Kingston,Canada,1995).\\
$^{~7}$~W.~B.~Bonnor, {\it  Gen.~Rel.~Grav.} {\bf 21}, 1143 (1989).\\
$^{~8}$~J.~R.~Gott III and M.~J.~Rees, {\it Mon.~Not.~Roy.~Astr.~Soc.} {\bf 227}, 453 (1987).\\
$^{~9}$~E.~W.~Kolb, {\it Astrophys.~J.} {\bf 344}, 543 (1989).\\
$^{10}$~P.~S.~Wesson, {\it Astrophys.~J.} {\bf 378}, 466 (1991).\\
$^{11}$~P.~S.~Wesson, {\it Physics Essays} {\bf 5}, 561 (1992).\\
$^{12}$~J.~Ponce de Leon, {\it Gen.~Rel.~Grav.} {\bf 25}, 1123 (1993).\\
$^{13}$~J.~M.~Overduin and F.~I.~Cooperstock, {\it  Phys.~Rev.} {\bf D58}, 43506 (1998).\\
$^{14}$~J.~M.~Overduin, {\it Astrophys.~J.} {\bf 517}, L1 (1999).\\
$^{15}$~S.~Leonard and K.~Lake, {\it  Astrophys.~J.} {\bf 441}, L55 (1995).\\
$^{16}$~G.~Abolghasem, A.~A.~Coley and D.~J.~Mc Manus, {\it J.~Math.~Phys.} {\bf 37}, 361 (1996).\\
$^{17}$~A.~Billyard and P.~S.~Wesson, {\it Phys.~Rev.} {\bf D53}, 731 (1996).\\
$^{18}$~M.~Mashhoon, P.~S.~Wesson and H.~Liu,{\it Gen.~Rel.~Grav.} {\bf 30}, 555 (1998).\\
$^{19}$~P.~S.~Wesson, B.~Mashhoon, H.~Liu and W.~N.~Sajko, {\it Phys.~Lett.} {\bf B456}, 34 (1999).\\
$^{20}$~H.~Liu and B.~Mashhoon, {\bf gr-qc/0005079} (2000).\\
$^{21}$~S.~S.~Seahra and P.~S.~Wesson, submitted to {\it Gen.~Rel.~Grav.}\\
$^{22}$~A.~Banerjee, B.~K.~Bhui and S.~Chatterjee, {\it Astron. Astrophys.} {\bf 232}, 893 (1990).\\
$^{23}$~H.~Liu and P.~S.~Wesson, {\it Int.~J.~Mod.~Phys.} {\bf D3}, 627 (1994).\\
$^{24}$~P.~S.~Wesson, H.~Liu and S.~S.~Seahra, {\it Astron.~Astrophys.} {\bf 358}, 425 (2000).\\
$^{25}$~P.~D.~Linde, {\it Inflation and Quantum Cosmology}(Academic Press, Boston,1999).\\
$^{26}$~F.~Hoyle, {\it Astrophys.~J.} {\bf 196}, 661 (1975).\\
\newpage
\begin{table}
\begin{centering}
\begin{tabular}{|c|c|c|} \hline
{}&$F<0$ & $F>0$\\ \hline\hline $\kappa>0$ and $R^A{}_{BCD}\neq 0$
in general&no big bang&$a(t,l)$ is complex\\ \hline $\kappa=0$,
$R^A{}_{BCD}=0$&one big bang&$a(t,l)$ is complex\\ \hline
$\kappa<0$ and $R^A{}_{BCD}\neq 0$ in general&two big bangs&big
bang and big crunch\\ \hline
\end{tabular}
\caption{Characteristics of the 5D Manifold and the 4D Cosmologies
embedded on the $\Sigma_l$ hypersurfaces}\label{table}
\end{centering}
\end{table}

\end{document}